\documentclass{llncs}

\usepackage{amssymb}
\usepackage{graphicx}
\usepackage{amsmath}
\usepackage[ruled,algo2e,linesnumbered,algonl]{algorithm2e}
\usepackage[free-standing-units,group-separator={,},binary-units=true]{siunitx}

\usepackage{color}

\newcommand{\vv}[1]{\mathbf{#1}}
\usepackage[colorlinks = true,
            linkcolor = blue,
            urlcolor  = blue,
            citecolor = blue,
            anchorcolor = blue]{hyperref}
\usepackage{caption}
\usepackage{subcaption}            
\begin{document}

\begin{frontmatter}


\title{Shared memory pipelined Parareal}

\author{Daniel Ruprecht\inst{1,2}}
\institute{School of Mechanical Engineering, LS2 9JT, Leeds, United Kingdom
\email{d.ruprecht@leeds.ac.uk}
\and
Institute of Computational Science, Universit{\`a} della Svizzera italiana, Via Giuseppe Buffi 13, CH-6900 Lugano, Switzerland
}

\maketitle

\begin{abstract}
For the parallel-in-time integration method Parareal, pipelining can be used to hide some of the cost of the serial correction step and improve its efficiency.
The paper introduces a basic OpenMP implementation of pipelined Parareal and compares it to a standard MPI-based variant.
Both versions yield almost identical runtimes, but, depending on the compiler, the OpenMP variant consumes about $7$\% less energy and has a significantly smaller memory footprint.
However, its higher implementation complexity might make it difficult to use in legacy codes and in combination with spatial parallelisation.
\end{abstract}


\end{frontmatter}

\section{Introduction}
Computational science faces a variety of challenges stemming from the massive increase in parallelism in state-of-the-art high-performance computing systems.
One important requirement is the development of new and inherently parallel numerical methods.
 Parallel-in-time integration has been identified as a promising direction of research for the parallelisation of the solution of initial value problems~\cite{DongarraEtAl2014}.

Probably the most widely studied and used parallel-in-time method is Parareal~\cite{LionsEtAl2001}, but see M. Gander's overview for a discussion of a variety of other methods~\cite{Gander2015_Review}.
Parareal iterates between an expensive fine integrator run in parallel and a cheap coarse method which runs in serial and propagates corrections forward in time.
While the unavoidable serial part limits parallel efficiency according to Amdahl's law, some of its cost can be hidden by using a so-called \emph{pipelined} implementation~\cite{Aubanel2011,Minion2010}.
Pipelining reduces the effective cost of the serial correction step in Parareal and therefore improves speedup.
Even more optimisation is possible by using an event-based approach~\cite{BerryEtAl2012}, but this requires a suitable execution framework that is not available on all machines.

Pipelining happens automatically when implementing Parareal in MPI but it is not straightforward in OpenMP and so far no shared memory version of Parareal with pipelining has been described.
Previous studies almost exclusively used MPI to implement Parareal and the very few using OpenMP considered only the non-pipelined version~\cite{RuprechtKrause2014_DDM}.
However, using shared memory can have advantages, since it avoids e.g. the need to allocate buffers for message passing. 
The disadvantage is that naturally OpenMP is limited to a shared memory unit.
Since  convergence of Parareal tends to deteriorate if too many parallel time slices are computed~\cite{GanderVandewalle2007_SISC} and given the trend to ``fat'' compute nodes with large numbers of cores, shared memory Parareal might nevertheless be an attractive choice.
It could be useful, e.g., for simulations of power grids and other applications where comparatively small systems of differential(-algebraic) equations have to be solved faster than real-time~\cite{LecouvezEtAl2016}.
This paper introduces an OpenMP-based version of pipelined Parareal and compares it to a standard MPI-based implementation.
It relies on standard features like parallelised loops, following a fork-join paradigm, while leaving the investigation of more recent OpenMP features providing task-base parallelism~\cite{openmp_tasks} for future work.

\section{The Parareal parallel-in-time integration method}
The starting point for Parareal is an initial value problem
\begin{equation}
	\label{eq:ivp}
	\dot{\vv{q}} = \mathbf{f} (\mathbf{q}(t),t), \ \vv{q}(0) = \vv{q}_{0}, \ t \in [0, T],
\end{equation}
which, in the numerical example below, arises from the spatial discretisation of a PDE ("method-of-lines") with $\vv{q} \in \mathbb{R}^{N_{\rm dof}}$ being a vector containing all degrees-of-freedom.
Let $\mathcal{F}_{\delta t}$ denote a numerical procedure for the approximate solution of \eqref{eq:ivp}, for example a Runge-Kutta method. 
Denote further by $\vv{q} = \mathcal{F}_{\delta t}\left( \vv{\tilde{q}}, t_{2}, t_{1} \right)$ the result of approximately integrating \eqref{eq:ivp} forward in time from some starting value $\mathbf{\tilde{q}}$ at a time $t_{1}$ to a time $t_{2} > t_{1}$ using $\mathcal{F}_{\delta t}$.

To parallelise the numerical integration of~\eqref{eq:ivp}, decompose $[0,T]$ into time-slices $[t_{p}, t_{p+1}]$, $p=0, \ldots, P-1$ where $P$ is the number of cores, equal to the number of processes (MPI) or threads (OpenMP).
For simplicity, assume here that all time slices have the same length and that the whole interval $[0,T]$ is covered with a single set of time slices so that no restarting or ``sliding window'' is required~\cite{SchreiberEtAl2015}.

Parareal needs a second time integrator denoted $\mathcal{G}_{\Delta t}$, which has to be cheap to compute but can be much less accurate (commonly referred to as the ''coarse propagator'').
It begins with a prediction step, computing rough guesses of the starting value $\vv{q}^0_{p}$ for each time slice by running the coarse method once.
Here, subscript $p$ indicates an approximation of the solution at time $t_p$.
It then computes the iteration 
\begin{equation}
	\label{eq:parareal}
	\vv{q}^{k}_{p+1} = \mathcal{G}_{\Delta t}(\vv{q}^{k}_{p}, t_{p+1}, t_{p}) + \mathcal{F}_{\delta t}(\vv{q}^{k-1}_{p}, t_{p+1}, t_{p}) - \mathcal{G}_{\Delta t}(\vv{q}^{k-1}_{p}, t_{p+1}, t_{p}) 
\end{equation}
concurrently on each time slice for $p=0,\ldots,P-1$, $k=1,\ldots,K$.
Because the computationally expensive evaluation of the fine propagator can be parallelised across time slices, iteration~\eqref{eq:parareal} can run in less wall clock time than running $\mathcal{F}_{\delta t}$ serially -- provided the coarse method is cheap enough and the number of required iterations $K$ is small.

The expected performance of Parareal can be described by a simple theoretical model~\cite{Minion2010}.
Denote by $c_{\rm c}$ the cost of integrating over one time slice using $\mathcal{G}_{\Delta t}$ and by $c_{\rm f}$ the cost when using $\mathcal{F}_{\delta t}$.
Because all time slices are assumed to consist of the same number of steps and an explicit method is used here, it can be assumed that $c_{\rm f}$ and $c_{\rm c}$ are identical for all time slices.
Neglecting overhead, speedup of Parareal using $K$ iterations against running the fine method in serial is approximately
\begin{equation}
	\label{eq:speedup_nopipe}
	s_{\rm np}(P) = \frac{ P c_{\rm f} }{ \left(1 + K \right) P c_{\rm c} + K c_{\rm f}} = \frac{1}{ \left(1 + K \right) \frac{c_{\rm c}}{c_{\rm f}} + \frac{K}{P}}.
\end{equation}

\begin{figure}[t]
	\centering
	\begin{minipage}[t]{0.4\textwidth}
		\includegraphics[width=0.99\textwidth]{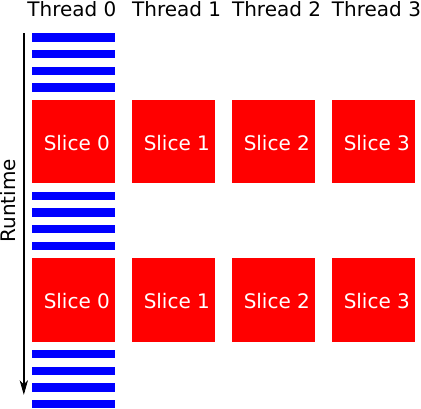}
	\end{minipage}
	\begin{minipage}[t]{0.4\textwidth}
		\includegraphics[width=0.99\textwidth]{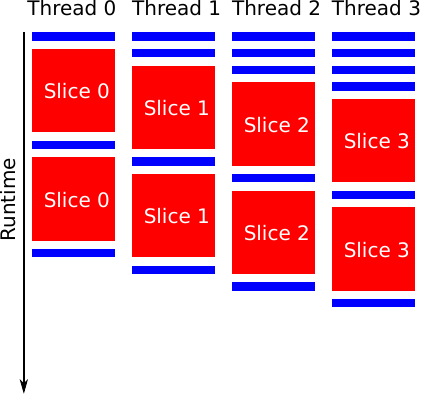}
	\end{minipage}
	\caption{Execution diagram for Parareal without (left) and with (right) pipelining. Red blocks correspond to $c_{\text{f}}$, the time needed to run the fine integrator $\mathcal{F}_{\delta t}$ over one time slice $[t_{p}, t_{p+1}]$ while blue blocks correspond to $c_{\text{c}}$, the time required by the coarse method $\mathcal{G}_{\Delta t}$. Pipelining allows to hide some of the cost of the coarse method. While it comes naturally when using MPI to implement Parareal (there, \texttt{Thread} would refer to a \texttt{Process}), using simple loop-based parallelism with OpenMP results in the non-pipelined version shown on the left.}	
	\label{fig:pipelining}	
\end{figure}

It is possible to hide some of the cost of the coarse propagator and the name \emph{pipelining} has been coined for this approach~\cite{Minion2010}.
Figure~\ref{fig:pipelining} sketches the execution diagrams of both a non-pipelined (left) and pipelined (right) implementation for four time slices.
As can be seen, pipelining reduces the effective cost of the coarse correction step in each iteration from $P \times c_{\textrm{c}}$ to $c_{\textrm{c}}$ -- but note that the initial prediction step still has cost $P \times c_{\rm c}$ as before.
For pipelined Parareal, estimate~\eqref{eq:speedup_nopipe} changes to
\begin{equation}
	\label{eq:speedup_pipe}
	s_{\rm p}(P) = \frac{P c_{\rm f}}{P c_{\rm c} + K c_{\rm c} + K c_{\rm f}} = \frac{1}{\left(1 + \frac{K}{P} \right) \frac{c_{\rm c}}{c_{\rm f}} + \frac{K}{P}}.
\end{equation}
Because $K/P \ll K$, the pipelined version allows for better speedup, that is $s_{\rm np}(P) \leq s_{\rm p}(P)$.
However, because pipelining only hides cost from the coarse integrator, the effect is smaller when the coarse method is very cheap and $c_{\rm c}/c_{\rm s} \ll 1$. In that case, the term $K/P$ dominates the estimate which is not affected by pipelining.

\section{Pipelined Parareal in OpenMP}\label{sec:implementation}
The implementation of Parareal with pipelining in OpenMP is sketched in Algorithm~\ref{alg:parareal_openmp_pipe}.
A description of how Parareal is implemented using MPI is available in the literature~\cite{ArteagaEtAl2015} and is therefore not repeated here.

Threads are spawned by the \texttt{OMP PARALLEL} directive in line~\ref{alg:openmp-pipe-parastart} and terminated by \texttt{OMP END PARALLEL} in line~\ref{alg:openmp-pipe-paraend}.
Manual synchronisation is required so that $P$ OpenMP locks are created using \texttt{OMP\_INIT\_LOCK} in line~\ref{alg:openmp-lock}, one for each thread.
During the fine integrator and update step, these locks are set and unset using \texttt{OMP\_SET\_LOCK} and \texttt{OMP\_UNSET\_LOCK} to protect buffers during writes and avoid race conditions.
\begin{algorithm2e}[t!]
        \caption{Parareal with pipelining using OpenMP}\label{alg:parareal_openmp_pipe}
        \SetKwInOut{Input}{input}
        \Input{Initial value $q_0$; number of iterations $K$}
        \SetCommentSty{textit}
        \DontPrintSemicolon       
	$q(0) \leftarrow q_0$\\
	$\texttt{OMP PARALLEL}$\label{alg:openmp-pipe-parastart}\\
	$P = \texttt{OMP\_GET\_MAX\_THREADS}()$\\
	$\texttt{OMP\_INIT\_LOCK}(P)$\label{alg:openmp-lock}\\
	$\texttt{OMP DO}$\label{alg:openmp-pipe-predict-start}\\
	\For{$p=0,P-1$}{
		$q(p) \leftarrow q(0)$\\
		\If{Thread not first}{
			$q(p) \leftarrow \mathcal{G}(q(p), t_{p}, 0)$
		}
		$g_c(p) \leftarrow \mathcal{G}(q(p), t_{p+1}, t_{p})$\\
	}
	$\texttt{OMP END DO NOWAIT}$\label{alg:openmp-pipe-predict-end}\\ 
	\For{$k=1,K$}{\label{alg:openmp-pipe-parareal-start}
		$\texttt{OMP DO ORDERED}$\label{alg:openmp-pipe-ordered-do}\\
		\For{$p=0,P-1$}{
			$\texttt{OMP\_SET\_LOCK(p)}$\label{alg:openmp-pipe-fine-start}\\
			$q(p) \leftarrow \mathcal{F}_{\delta t}(q(p), t_{p+1}, t_p)$\\
			$\delta q(p) \leftarrow q(p) - q_c(p)$\\
			$\texttt{OMP\_UNSET\_LOCK(p)}$\label{alg:openmp-pipe-fine-end}\\
			$\texttt{OMP ORDERED}$\label{alg:openmp-pipe-correction-start}\\
			\If{Thread is first}{
				$\texttt{OMP\_SET\_LOCK(0)}$\\
				$q(0) \leftarrow q_0$\\
				$\texttt{OMP\_UNSET\_LOCK(0)}$\\
			}
			$q_c(p) \leftarrow \mathcal{G}_{\Delta t}(q(p), t_{p+1}, t_{p})$\\
			\If{Thread not last}{
				$\texttt{OMP\_SET\_LOCK(p+1)}$\\
				$q(p+1) \leftarrow q_c(p) + \delta q(p)$\label{alg:openmp-pipe-correction}\\
				$\texttt{OMP\_UNSET\_LOCK(p+1)}$\\
			}
			$\texttt{OMP END ORDERED}\label{alg:openmp-pipe-correction-end}$
		}
		$\texttt{OMP END DO NOWAIT}$\label{alg:openmp-pipe-last-no-wait}
	}\label{alg:openmp-pipe-parareal-end}
	$\texttt{OMP END PARALLEL}$\label{alg:openmp-pipe-paraend}
\end{algorithm2e}

The algorithm consists of the following parts:
\begin{itemize}
	\item \textbf{Prediction step:} lines~\ref{alg:openmp-pipe-predict-start}--\ref{alg:openmp-pipe-predict-end}.
Each thread is computing its own coarse prediction of its starting value $\vv{q}^{0}_{p}$ in a parallelised loop.
The coarse value $\mathcal{G}_{\Delta t}(\vv{q}^{0}_p, t_{p+1}, t_p)$ is also computed and stored for use in the first iteration.
The later the time slice (indicated by a higher thread number $p$), the more steps the thread must compute and thus the larger its workload (cf. Figure~\ref{fig:pipelining}).
Therefore, at the end of the coarse prediction loop, the \texttt{NOWAIT} clause is required to avoid implicit synchronisation and enable pipelining.
	\item \textbf{Parareal iteration:} lines~\ref{alg:openmp-pipe-parareal-start}--\ref{alg:openmp-pipe-parareal-end}.
Here, both the fine integrator and update step are performed inside a single loop over all time slices, parallelised by \texttt{OMP DO} directives.
Because parts of the loop (the update step) have to be executed in serialised order, the \texttt{ORDERERD} directive has to be used in line~\ref{alg:openmp-pipe-ordered-do}.
Again, to avoid implicit synchronisation at the end of the loop, the \texttt{NOWAIT} clause is required in line~\ref{alg:openmp-pipe-last-no-wait}.
		\begin{itemize}
			\item \textbf{Fine integrator:} lines~\ref{alg:openmp-pipe-fine-start}--\ref{alg:openmp-pipe-fine-end}. Before the fine integrator is executed, an \texttt{OMP\_LOCK} is set to indicate that the thread will start writing into buffer $q(p)$. 
Because thread $p-1$ accesses this buffer in its update step, locks are necessary to prevent race conditions and incorrect solutions.
After the lock is set, the thread proceeds with the computation of $\mathcal{F}_{\delta t}(\vv{q}^k_p, t_{p+1}, t_p)$ and computation of the difference $\delta q$ between coarse and fine value.
Then, since $q(p)$ is now up to date and $\delta q$ ready, the lock can be released.
		\item \textbf{Update step:} lines~\ref{alg:openmp-pipe-correction-start}--\ref{alg:openmp-pipe-correction-end}. 
The update step has to be performed in proper order, from first to last time slice. 
Therefore, it is enclosed in \texttt{ORDERED} directives, indicating that this part of the loop is to be executed in serial order.
Then, as in the two other versions, the update step is initialised with $\vv{q}^{k+1}_0 = \vv{q}_0$.
For every time slice, the coarse value of the updated initial guess is computed and the update performed.
The updated end value is written into buffer $q(p+1)$ to serve as the new starting value for the following time slice.
However, to prevent thread $p$ from writing into $q(p+1)$ while thread $p+1$ is still running the fine integrator, thread $p$ sets \texttt{OMP\_LOCK} number $p+1$ while performing the update.
		\end{itemize}
\end{itemize}

The implementation described here computes a fixed number of iterations $K$.
While useful for testing, this is not necessarily optimal since it will perform iterations even on time slices that have already converged.
A simple optimisation would be to leave threads idle for time slices that have converged.
For larger problems where not the whole time interval can be covered with time slices,  either restarting or some form of ``sliding window'' should be employed.
Both cases require to replace the outer \texttt{FOR} loop by some form of adaptive control of iterations and are not considered here.

\section{Numerical results}\label{sec:results}
This section compares the pipelined OpenMP implementation to a straightforward MPI variant with respect to runtime, memory footprint and energy consumption.
The code used here for benchmarking is written in Fortran 90 and available for download~\cite{PararealF90}.
It is special-purpose and solves the nonlinear 3D Burgers' equation
\begin{equation}
	\label{eq:burger}
	u_{t} + u \cdot \nabla u = \nu \Delta u
\end{equation}
on $[0,1]^{3} \subset \mathbb{R}^{3}$ with periodic boundary conditions using finite differences.
Both implementations of Parareal use the same modules to provide the coarse and fine integrator and spatial discretisation.
Tests guarantee that the two implementations of Parareal produce results that are identical up to a tolerance of $\varepsilon = 10^{-14}$ and thus essentially to round-off error.
To detect possible race conditions, the comparison is run multiple times. 
Up to 100 instances of the test were passed on both used architectures.
Furthermore, both implementations of Parareal use three auxiliary buffers per time-slice: $q$ to store the fine value and for communication, $\delta q$ to store the difference $\mathcal{F}_{\delta t}(q) - \mathcal{G}_{\Delta t}(q)$ needed in the correction step and $q_c$ to store the coarse value from the previous iteration.

A strong stability preserving Runge-Kutta method (RK3-SSP)~\cite{ShuOsher1989} with a fifth order WENO finite difference discretisation~\cite{ShuOsher1989} for the advection term and a fourth order centred difference for the diffusion term is used for $\mathcal{F}_{\delta t}$.
For $\mathcal{G}_{\Delta t}$, a first order forward Euler with a simple first order upwind stencil for the advection term and a second order centred stencil for the diffusive term is used.
Being a simplified version of the Navier-Stokes equations,~\eqref{eq:burger} is a popular benchmark and finite difference stencils are a widely used motif in computational science, so that the results can be expected to hold for more general scenarios, at least qualitatively.

Parameters for the simulation are a viscosity parameter of $\nu = 0.02$ and a spatial discretisation on both levels with $N_x = N_y = N_z = 40$ grid points in every direction.
The simulation is run until $T=1.0$ with a coarse time step of $\Delta t = 1/192$ and a fine step of $\delta t = 1/240$.
Because of the high computational cost of the WENO-5 method in comparison to a cheap first order upwind scheme and the fact that RK3SSP needs three evaluations of the right hand side per step while the Euler method needs only one, the coarse propagator is about a factor of forty faster, despite the fact that the coarse step is only a factor of $1.25$ larger than the fine.
Using a coarse time step $\Delta t \approx \delta t$ prevents stability issues in the coarse propagator and improves convergence of Parareal.

To fix the number of iterations to a meaningful value which guarantees comparable accuracy from Parareal and serial fine integrator, we estimate the discretisation error of $\mathcal{F}_{\delta t}$ by comparing against a reference solution with time step $\delta t / 10$.
This gives estimates for the fine relative error at $T=1$ of about $e_{\rm fine} \approx 5.9 \times 10^{-5}$ and for the coarse error of about $e_{\rm coarse} \approx 7.3 \times 10^{-2}$.
For $P=24$ time slices, after three iterations, the defect between Parareal and the fine solution is approximately $1.4 \times 10^{-4}$, after four iterations $1.5 \times 10^{-5}$.
We therefore fix the number of iterations to $K=4$ so that for all values of $P$ Parareal produces a solution with the same accuracy as the fine integrator.

Benchmarks are run on two systems.
The first is a commodity work station with an 8-core Intel Xeon(R) E5-1660 CPU and 32 GigaByte of memory running CentOS Linux 7.
Flags \texttt{-O3} and  \texttt{-fopenmp} were used when compiling the code for maximum optimisation and to enable OpenMP.
As the code is stand alone no external libraries have to be linked.
The used MPI implementation is mpich-3.0.4, compiled with GCC-4.8.5.

The second system is one node of \textsc{Piz Dora} at the Swiss National Supercomputing Centre CSCS.\footnote{\url{http://www.cscs.ch/}}
\textsc{Dora} is a Cray XC40 with a total of $1,256$ compute nodes.
Each node contains two $12$-core Intel Broadwell CPUs and has 64 or 128 GigaByte of RAM and nodes are connected through a Cray Aries interconnect, using a dragonfly network topology.
Two compilers are tested, the GCC-4.9.2 and the Cray Fortran compiler version 8.3.12.
Both use the MPICH MPI library version 7.2.2.
Compiler flags \texttt{-O3} and \texttt{-fopenmp} (GCC) or \texttt{omp} (Cray compiler) are used for compilation.
Performance data for each completed job is generated using the Cray \emph{Resource Utilisation Reporting} tool RUR~\cite{Rur2013}.
RUR collects compute node statistics for each job and provides data on user and system time, maximum memory used, amount of I/O operations, consumed energy and other metrics.
However, it only collects data for a full node and not for individual CPUs or cores.

\subsection{Wall clock time and speedup}\label{subsec:runtime}
At first, runtime and speedup compared to the serial execution of the fine integrator are assessed.
On both the Linux work station and \textsc{Dora}, five runs are performed for each variant of Parareal and each value of $P$ and the average runtime is reported.
Measured runtimes are quite stable across different runs: the largest relative standard deviation of all performed five-run ensembles is smaller than $0.05$ on Linux and smaller than $0.01$ on \textsc{dora}.
Therefore, plots show only the average values without error bars, because those are hardly recognisable and clutter the figure.

\begin{figure}[t]
	\centering
	\begin{subfigure}[t]{0.49\textwidth}
		\centering
		\caption{\textsc{Piz Dora}}\label{fig:runtime_dora_cray}
		\includegraphics[scale=1]{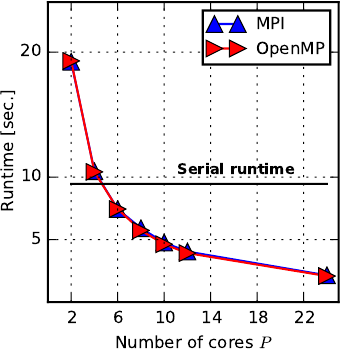}
	\end{subfigure}
	\begin{subfigure}[t]{0.49\textwidth}
		\centering
		\caption{\textsc{Linux work station}}\label{fig:runtime_cub}
		\includegraphics[scale=1]{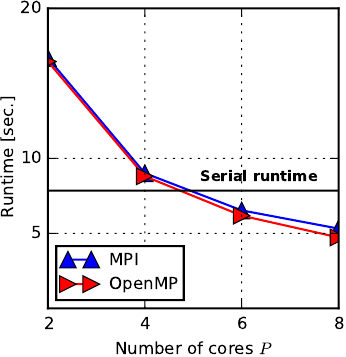}
	\end{subfigure}
    \caption{Five run averages of runtime with relative standard deviation below $0.01$ on \textsc{dora} (\ref{fig:runtime_dora_cray}) and $0.05$ on a Linux work station (\ref{fig:runtime_cub}).}
        \label{fig:runtimes}      
\end{figure}

\begin{figure}[t!]
	\centering
	\begin{subfigure}[t]{0.49\textwidth}
		\centering
		\caption{\textsc{Piz Dora (GCC)}}\label{fig:runtime_dora_gcc}
		\includegraphics[scale=1]{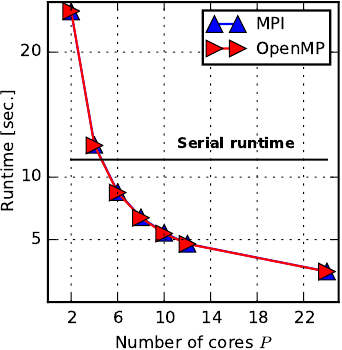}
	\end{subfigure}
	\begin{subfigure}[t]{0.49\textwidth}
		\centering
		\caption{\textsc{Piz Dora (GCC)}}\label{fig:speedup_dora_gcc}
		\includegraphics[scale=1]{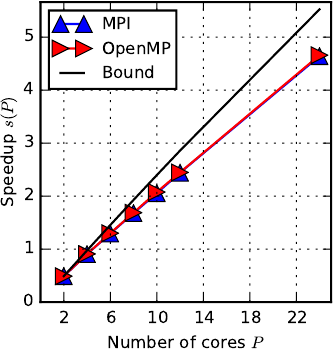}
	\end{subfigure}
    \caption{Runtime (\ref{fig:runtime_dora_gcc}) and speedup (\ref{fig:speedup_dora_gcc}) on Piz Dora using GCC.}
        \label{fig:dora_gcc}        
\end{figure}

Figure~\ref{fig:runtimes} shows runtimes in seconds depending on the number of cores on \textsc{Dora} using the Cray compiler (left) and Linux (right).
The Cray compiler generates faster code than GCC in the case studied here, but for comparison results using GCC on \textsc{Dora} are shown in Figure~\ref{fig:dora_gcc}.
The runtime of the serial fine integrator is indicated by a horizontal black line.
The CPU in the Linux system has a slightly higher clock speed so that runtimes are a bit faster than on \textsc{dora}. 
For $P=8$ cores, for example, OpenMP-Parareal runs in slightly less than \SI{5}{\second} there while taking about \SI{5.7}{\second} on \textsc{dora}.
Differences between the OpenMP and MPI version are small on both systems, but for $P=8$ OpenMP-Parareal is marginally faster than the MPI version on the work station.

\begin{figure}[t!]
	\centering
	\begin{subfigure}[t]{0.49\textwidth}
		\centering
		\caption{\textsc{Piz Dora}}\label{fig:speedup_dora_cray}
		\includegraphics[scale=1]{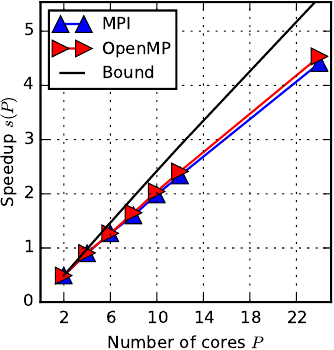}
	\end{subfigure}
	\begin{subfigure}[t]{0.49\textwidth}
		\centering
		\caption{\textsc{Linux work station}}\label{fig:speedup_cub}
		\includegraphics[scale=1]{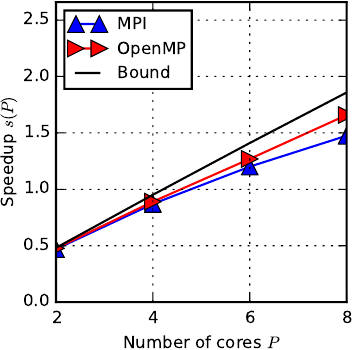}
	\end{subfigure}
    \caption{Speedup computed from average runtimes shown in Figure~\ref{fig:runtimes} on \textsc{dora} (\ref{fig:speedup_dora_cray}) and a Linux work station (\ref{fig:speedup_cub}).}
        \label{fig:speedup}        
\end{figure}

In addition, Figure~\ref{fig:speedup} shows the speedup relative to the fine integrator run serially.
The black line indicates projected speedup according to~\eqref{eq:speedup_pipe}.
Both versions fall short of the theoretically possible speedup. 
Because of overheads, running $P$ instances of $\mathcal{F}_{\delta t}$ on $P$ cores does take longer than running a single instance.
However, differences between MPI and OpenMP are small.
As far as runtimes and speedup are concerned, there is no indication that using the more complex OpenMP version provides significant benefits.

\subsection{Memory footprint}\label{subsec:memory}
\begin{figure}[t!]
	\centering
	\begin{subfigure}[t]{0.49\textwidth}
		\centering
		\caption{GCC compiler}
		\label{fig:memory_gcc}
		\includegraphics[scale=1.0]{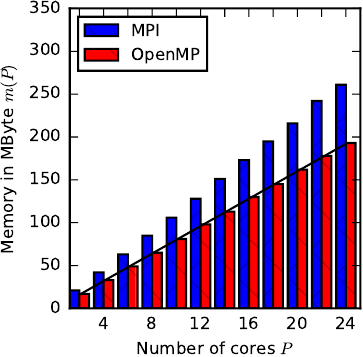}	
	\end{subfigure}
	\hfill
	\begin{subfigure}[t]{0.49\textwidth}
		\centering
		\caption{Cray compiler}
		\label{fig:memory_cray}
		\includegraphics[scale=1.0]{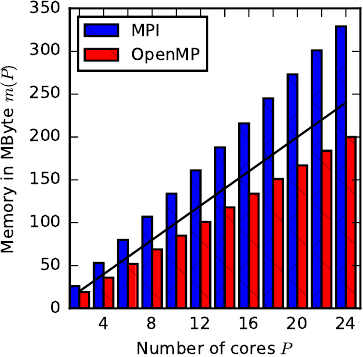}
	\end{subfigure}
	\caption{Maximum memory allocated in MegaByte for GCC (\ref{fig:memory_gcc}) and Cray compiler (\ref{fig:memory_cray}) for the three different versions of Parareal depending on the number of used cores $P$. The black line indicates expected memory consumption computed as number of cores time memory footprint of serial fine integrator.}
	\label{fig:memory}
\end{figure}

The memory footprint of the code is measured only on \textsc{dora} where RUR is available.
In contrast to runtime and energy, the memory footprint, as expected, does not vary between runs. 
Therefore~Figure~\ref{fig:memory} shows a visualisation of the data from a single run with no averaging.
The bars indicate the maximum required memory in MegaByte (\SI{}{\mega\byte}) while the black line indicates the expected memory consumption using $P$ cores computed as
\begin{equation}
	\label{eq:memory}
	m(P) = P \times m_{\rm serial}
\end{equation}
where $m_{\rm serial}$ is the value measured for a reference run of the fine integrator.
Because copies of the solution have to be stored for every time slice, the total memory required for Parareal can be expected to increase linearly with the number of cores in time.
Note, however, that memory required \emph{per core} stays constant if it follows~\eqref{eq:memory}.

For the OpenMP variant compiled with GCC, the memory footprint shown in Figure~\ref{fig:memory_gcc} exactly matches the expected values.
The Cray compiler, shown in Figure~\ref{fig:memory_cray}, leads to a smaller than expected memory footprint, but memory requirements still increase linearly with the number of time slices.
For both compilers, the MPI version causes a noticeable overhead in terms of memory footprint, most likely because of internal allocation of additional buffers for sending and receiving~\cite{Rabenseifner2009}.
For both OpenMP and MPI, the total memory footprint is \emph{larger} for the Cray than for the GCC compiler, but the effect is much more pronounced for the MPI implementation (\SI{329}{\mega\byte} versus \si{261}{\mega\byte}) than for the OpenMP variant (\si{200}{\mega\byte} versus \si{193}{\mega\byte}).

It is important to note that both implementations allocate three auxiliary buffers per core.
The overhead in terms of memory in MPI does thus not simply stem from allocating an additional buffer for communication but comes from within the MPI library.
The OpenMP implementation avoids this overhead.

It should be pointed out that the MPI baseline relies on two-sided \texttt{MPI\_RECV} and \texttt{MPI\_SEND} directives for communications.
Exploring whether an implementation based on one-sided remote memory access in MPI~\cite{mpi_rma} retains the advantages of OpenMP would be an interesting continuation of the presented work.

\subsection{Energy-to-solution}\label{subsec:energy}
%
The RUR tool reports the energy-to-solution for every completed job.
Because RUR can only measure energy usage for a full node, results are reported here for runs using the full number of $P=24$ cores available on a \textsc{dora} node.
Unfortunately, this makes measuring the energy consumption of the fine propagator largely meaningless since it can use only a single core of the node.
Therefore, no corresponding measurements were taken to quantify the overhead of Parareal in terms of energy.

In contrast to runtimes and memory footprint, energy measurements show significant variations between runs due to random fluctuations.
Thus, the presented values are averages over ensembles of $50$ runs for each version of Parareal.
This number of runs has been sufficient to reduce the relative standard deviation to below $0.09$ in both configurations and therefore gives a robust indication of actual energy requirements.
Table~\ref{tab:energy} shows the results including $95\%$ confidence intervals, assuming energy-to-solution is normally distributed.
For comparison, energy-to-solution is also shown for a simple non-pipelined OpenMP implementation where only the loop around $\mathcal{F}_{\delta t}$ is parallelised using \texttt{OMP PARALLEL DO}.

\begin{table}[t!]
\centering
\begin{tabular}{|c|ccc|} \hline
		        \multicolumn{4}{|c|}{\textbf{GCC compiler}} \\ \hline
		       &  $\varnothing$energy (Joule)  & confidence (Joule) &  $\varnothing$runtime (sec.) \\ \hline
MPI                 & 844.04                                & $\pm$~15.89 & 2.455  \\
OpenMP         & 801.14                                     & $\pm$~13.18  & 2.930  \\
OpenMP(pipelined) & 783.72                                       & $\pm$~11.53 & 2.448  \\ \hline 
\end{tabular}
\begin{tabular}{|c|ccc|} \hline
		        \multicolumn{4}{|c|}{\textbf{Cray compiler}} \\ \hline
		       &  $\varnothing$energy (Joule)  & confidence (Joule) &  $\varnothing$runtime (sec.) \\ \hline
MPI                 & 784.24                              & $\pm$~11.93  & 2.146    \\
OpenMP         &  833.12                                    &  $\pm$~20.99 &  2.400 \\
OpenMP(pipelined) & 784.72                                     &  $\pm$~12.18 &  2.088   \\ \hline
\end{tabular}
\caption{Average energy-to-solution for the three different variants run on $P=24$ cores. Since RUR only measures energy consumption of a full node, it was not possible to meaningully measure energy-to-solution of the serial propagator and quantify the energy overhead.}
\label{tab:energy}
\end{table}
When using the GCC compiler, MPI-Parareal consumes more energy than both the pipelined and non-pipelined OpenMP versions.
The averages are well outside the confidence interval for the MPI version, so this is very unlikely just a chance result.
Moreover, because runtimes are almost identical for MPI and pipelined OpenMP, the differences in energy-to-solution cannot solely be attributed to differences in time-to-solution.
This is supported by the fact that non-pipelined OpenMP, despite being significantly slower, still consumes less energy than the MPI variant.
Tracking down the precise reason for the differences in energy-to-solution and power requirement will require detailed tracing of power uptake which is only possible on specially prepared machines~\cite{Canturk2003} and thus left for future work.

For code generated with the Cray compiler, both OpenMP and MPI lead to almost identical energy requirements.
Confidence intervals are $784.24 \pm 11.93$\si{}{\joule} for MPI and $784.24 \pm 12.18$\si{}{\joule} for OpenMP.
It seems likely that the compiler optimises the message passing to take advantage of the shared memory on the single node.
Supposedly, the MPI version handles communication in a way that is similar to what is explicitly coded in the OpenMP version.
However, as shown in Subsection~\ref{subsec:memory}, this automatic optimisation comes at the expense of a significantly larger memory footprint.

Note that the energy consumption of Parareal has previously been studied~\cite{ArteagaEtAl2015}.
By comparing against a simple theoretical model, it has been shown that the energy overhead of Parareal (defined as energy-to-solution of Parareal divided by energy-to-solution of the fine serial integrator), is mostly due to Parareal's intrinsic suboptimal parallel efficiency.
While improving parallel efficiency of parallel-in-time integration clearly remains the main avenue for improving energy efficiency, the results here suggest that in some cases a shared memory approach can provide non-trivial additional savings.

\section{Summary}
The paper introduces and analyses an OpenMP implementation of the parallel-in-time method Parareal with pipelining.
Pipelining allows to hide some of the cost of the serial coarse correction step in Parareal and is important to optimise its efficiency (even though it cannot relax the inherent limit on parallel efficiency given by the inverse of the number of required iterations).
Pipelining comes naturally in a distributed memory MPI implementation, but is not straightforward when using OpenMP.
The new OpenMP implementation is compared to the standard MPI variant in terms of runtime, memory footprint and energy consumption for both a Cray compiler and the GCC.
Both versions produce essentially identical runtimes.
For both compilers, using OpenMP leads to reductions in memory footprint, but the effect is more pronounced for the Cray compiler.
In terms of energy-to-solution, the results strongly depend on the compiler: while for GCC the OpenMP version is more energy efficient than the MPI version, there is no difference for the Cray compiler.

The results show that contemplating a shared memory strategy to implement ``parallel-across-the-steps'' methods like Parareal can be worthwhile.
Even though it is more complicated, it can reduce memory requirements.
However, more advanced features like task-based parallelism in OpenMP, remote memory access for MPI or a more advanced iteration control to prevent superfluous computation on converged time slices are not explored here.
Another potential caveat is whether the benefits carry over to the full space-time parallel case, where a parallel-in-time method is combined with spatial decomposition.
For Parareal without pipelining the potential of such a hybrid space-time parallel approach has been illustrated~\cite{RuprechtKrause2014_DDM} but whether this applies to the pipelined version introduced here remains to be seen.


\bibliographystyle{splncs03}
\bibliography{pint,hpc_mod,MyCodes,other}







\end{document}